\documentclass[twocolumn,preprintnumbers,amsmath,aps]{revtex4}
\usepackage[dvips]{graphicx}
\usepackage{dcolumn}
\usepackage{bm}
\usepackage{color}
\usepackage{multirow}
\usepackage{amsmath}
\begin {document}
\def\eq#1{(\ref{#1})}
\def\fig#1{Fig.\hspace{1mm}\ref{#1}}

\newcommand{\kvec}{\mbox{{\scriptsize {\bf k}}}}

\newcommand{\qvec}{\mbox{{\scriptsize {\bf q}}}}

\title{
Unbalanced superconductivity induced by the constant electron-phonon coupling\\ on a square lattice}
\author{Kamila A. Szewczyk$^{\left(1\right)}$}
\email{kamila.szewczyk@ajd.czest.pl}
\author{Rados{\l}aw Szcz{\c{e}}{\'s}niak$^{\left(1, 2\right)}$}
\author{Dominik Szcz{\c{e}}{\'s}niak$^{\left(1\right)}$}
\affiliation{$^1$ Institute of Physics, Jan D{\l}ugosz University in Cz{\c{e}}stochowa, Ave. Armii Krajowej 13/15, 42-200 Cz{\c{e}}stochowa, Poland}
\affiliation{$^2$ Institute of Physics, Cz{\c{e}}stochowa University of Technology, Ave. Armii Krajowej 19, 42-200 Cz{\c{e}}stochowa, Poland}
\date{\today} 
\begin{abstract}
In the present paper, we analyze the properties of the unbalanced superconducting state on a square lattice with the constant value of the electron-phonon coupling function. We conduct our analysis in the framework of the Eliashberg formalism, explicitly considering {\bf k}-dependence of the electron and phonon dispersion relations. It is found that the balanced superconducting state does not induce itself in the system due to the high value of the electron effective mass. However, in the unbalanced case the thermodynamic properties of the superconducting condensate can be distinctly different from the predictions of the Bardeen-Cooper-Schieffer theory; when the coupling constant value in the diagonal channel of the self-energy is diminished comparing to the non-diagonal channel. This is due to the reduced dimensionality of the tested system and the strong-coupling effects included in the Eliashberg formalism.
\end{abstract}
\maketitle
\noindent{\bf Keywords:} unbalanced superconducting state, two-dimensional systems, thermodynamic properties\\
%

By the \textit{unbalanced superconducting state} we may understand the superconducting phase for which the value of the coupling constant in the diagonal channel of the self-energy is different from the value of the coupling constant in the Cooper (non-diagonal) channel. In this context, the unbalanced superconducting state may occurs in situations when, in addition to the linear electron-phonon coupling, another interaction is taken into account. For example, the superconducting state induced by the electron-phonon interaction and the spin-fluctuation (paramagnon) scattering has the effective coupling constant in the diagonal channel ($\lambda_{D}$) equal to: $\lambda_{ph}+\lambda_{sf}$ (for $s$-wave and $p$-wave symmetry), while in the non-diagonal channel we have: $\lambda_{ND}=\lambda_{ph}-\lambda_{sf}$ ($s$-wave symmetry) or $\lambda_{ND}=\lambda_{sf}$ ($p$-wave symmetry) \cite{Rietschel1979A, Rietschel1980A, Fay1980A, Daams1981A, Fay1982A, Leavens1983A}; the symbols  $\lambda_{ph}$ and $\lambda_{sf}$ represent the electron-phonon coupling constant and the coupling constant for spin fluctuations, respectively. Note that in the case under consideration $\lambda_{D}$ is always greater than $\lambda_{ND}$, what from a physical point of view corresponds to the anomalous growth of the electron effective mass ($m^{\star}_{e}\sim \left(1+\lambda_{D}\right)m_{e}$) and the drop in the critical temperature ($T_{c}$) value \cite{Carbotte1990A, Carbotte2003A}. Therefore, an effect which is not desirable from the viewpoint of the high-$T_{c}$ superconductivity.

Another case of the unbalanced superconductivity can be distinguished when the linear electron-phonon interaction is complemented by the electron-electron-phonon interaction resulting from the existence of the strong electron correlations modeled by the on-site Hubbard repulsion \cite{Szczesniak2012A, Szczesniak2017A}. The discussed model can be used to explain the anomalous dependence of the order parameter on the temperature and the origin of the pseudogap. It turns out that the electron-electron-phonon interaction also leads to the unbalance of the superconducting state. This unbalance has complex character because $\lambda_{D}$ depends on the average number of electrons on the site ($\left<n\right>$) according to the formula: $\alpha_{0}+\alpha_{1}\left<n\right>+\alpha_{2}\left<n\right>^{2}$, whereas $\lambda_{ND}$ has the form: $\alpha_{0}+\alpha_{1}\left<n\right>$, where 
$\alpha_{i}$ represents respectively defined coefficients \cite{Szczesniak2017A}. Note that the explicit dependence of the effective coupling constants $\lambda_{D}$ and $\lambda_{ND}$ on $\left<n\right>$ is the direct reason for the suppression of the superconducting state near the half-filled electron band; the term $\alpha_{2}\left<n\right>^{2}$ becomes dominant, due to the sufficiently large $\alpha_{2}$. It is worth emphasizing that this effect is observed in cuprates \cite{Bednorz1986A, Bednorz1988A}. Obviously, full understanding of the properties of the superconducting state in cuprates requires also consideration of the dependence of the coupling function on the wave vector, which leads to the ${\bf k}$-dependent order parametr \cite{Dagotto1994A, Szczesniak2014A}. For instance, in the framework of the Eliashberg formalism, the dependence of the coupling function on $\bf{k}$ can be modeled through the appropriate factorization of the spectral function \cite{Rieck1990A}: $\alpha^{2}F\left({\bf k}, {\bf k}',\omega\right)\simeq \alpha^{2}F_{s}\left(\omega\right)+\alpha^{2}F_{d}\left(\omega\right)\Psi_{d}\left({\bf k}\right)\Psi_{d}\left({\bf k}'\right)$, where the first term represents weak pairing in the $s$-wave channel, and the second one is the pairing in the $d$-wave channel - most likely resulting from the existence of the strong electron correlations. It can be shown that the inclusion of the anisotropic coupling function leads to the another unbalance of the superconducting state, wherein $\lambda_{ND}>\lambda_{D}$ due to the different angular average \cite{Rieck1990A, Jiang1993A, Zasadzinski2003A}.

To our knowledge, the literature of the subject is currently composed of only one work written by Cappelluti and Ummarino \cite{Cappelluti2007A}. Specifically, this study presents systematic analysis of the unbalanced coupling effects on the superconducting properties in the three-dimensional system with the half-filled electron band. However, therein the dependence of the order parameter on the wave vector {\bf k} is omitted, for the constant value of the electron-boson coupling function. In a result, the discussed analysis shows strong increase of the critical temperature value for $\lambda_{ND}>\lambda_{D}$. Cappelluti and Ummarino also predict the finite critical value of the electron-boson coupling, where the superconducting critical temperature diverges but the zero temperature gap is still finite. However, in the real systems the finite bandwidth effects remove the analytical divergence of $T_{C}$. Moreover, paper \cite{Cappelluti2007A} suggest that both the balanced and unbalanced superconducting state can be induced in the system, where the latter one presents low value of $T_{C}$ ($\lambda_{ND}<\lambda_{D}$).

In reference to the above, present paper is the next step towards systematic study of the properties of the unbalanced superconducting state. Specifically, as a case study we choose the superconducting state induced in the two-dimensional system (the square lattice); a particularly interesting case which can mimics behavior of the popular and perspective quasi-two-dimensional systems like cuprates \cite{Bednorz1986A, Bednorz1988A, Dagotto1994A}. In our analysis, we assume that the superconducting state is induced by the linear electron-phonon interaction \cite{Frohlich1950A, Frohlich1952A, Frohlich1954A}, similarly like in the work of Cappelluti and Ummarino \cite{Cappelluti2007A}. However, contrary to \cite{Cappelluti2007A}, we take into account explicit dependence of the order parameter on the wave vector {\bf k}, due to the already mentioned potential importance of {\bf k}-dependent coupling function in describing superconducting state in quasi-two-dimensional materials. Notably, the anisotropy of the order parameter arises here as a result of employing explicit forms of both the electronic and phonon dispersion relation for the tested system. In our opinion, such approach, hitherto not considered, may result in revealing essential qualitative differences when comparing to the present state of knowledge.

%

\vspace*{0.25cm}

In the present work, discussed study is employed within the Eliashberg equations, which can be derived only when analysis of the superconducting condensate is conducted with the second-order accuracy with respect to the electron-phonon coupling function ($g_{\kvec, \kvec+\qvec}$) \cite{Eliashberg1960A}. It should be emphasized that the Eliashberg formalism far exceed the mean-field Bardeen-Cooper-Schrieffer (BCS) theory \cite{Bardeen1957A, Bardeen1957A}, most importantly, due to the explicit incorporation of the strong-coupling effects in its approach \cite{Carbotte1990A, Carbotte2003A}. In practice, the Eliashberg equations compose the non-linear system of equations, which allows determination of the superconducting state order parameter ($\Delta_{\kvec}\left(i\omega_{n}\right)$), as well as, the wave function renormalization factor ($Z_{\kvec}\left(i\omega_{n}\right)$), the band energy shift function ($\chi_{\kvec}\left(i\omega_{n}\right)$), and the chemical potential ($\mu$). Herein, $\omega_{n}$ is the fermion Matsubara frequency given by the expression: $\omega_{n}=\pi/\beta\left(2n-1\right)$, where $\beta=1/k_{B}T$ with $k_{B}$ denoting the Boltzmann constant. We note that the absolute value of order parameter directly corresponds to the half-width of energy gap on the Fermi surface. Hereafter, the wave function renormalization factor determines the renormalization of the electron band mass ($m_{e}$) by the electron-phonon interaction. Finally, the $\chi_{\bf k}\left(i\omega_{n}\right)$ function renormalizes the electron band energy ($\varepsilon_{\kvec}$), where the electronic dispersion relation for the square lattice has the form: 
$\varepsilon_{\kvec}=-2t\left[\cos\left(k_{x}\right)+\cos\left(k_{y}\right)\right]$, by taking into account only nearest neighbor hopping integrals ($t$).

Herein, the starting point for our analysis is constituted by the following full set of the Eliashberg equations which characterize the unbalanced superconducting state ($\gamma$ determines the degree of unbalance):
\begin{widetext}
\begin{equation}
\label{r1}
\varphi _{\bf k} \left( i\omega _{n}\right) = \frac{1}{\beta N} \sum _{\omega _{m}{\bf q}}K _{\bf kq} \left( \omega _{n}-\omega _{m}\right) D_{{\bf k}-{\bf q}} ^{-1} \left( i\omega _{m}\right) \varphi _{{\bf k}-{\bf q}}\left( i\omega _{m}\right),
\end{equation}
\begin{equation}
\label{r2}
Z_{\bf k}\left( i\omega _{n}\right) = 1+ \frac{\gamma}{\beta \omega _{n} N}  \sum _{\omega _{m}{\bf q}}K _{\bf kq} \left( \omega _{n}-\omega _{m}\right) D_{{\bf k}-{\bf q}} ^{-1} \left( i\omega _{m}\right) \omega _{m} Z _{{\bf k}- {\bf q}} \left( i\omega _{m}\right),
\end{equation}
\begin{equation}
\label{r3}
\chi _{\bf k} \left( i\omega _{n}\right) = \frac{\gamma}{\beta N} \sum _{\omega _{m}{\bf q}}K _{\bf kq} \left( \omega _{n}-\omega _{m}\right) D_{{\bf k}-{\bf q}} ^{-1} \left( i\omega _{m}\right) \left[\varepsilon_{{\bf k}-{\bf q}}-\mu+ \chi _{{\bf k}-{\bf q}} \left( i\omega _{m}\right) \right],
\end{equation}
and
\begin{equation}
\label{r4}
\left \langle n  \right \rangle = 1- \frac{2}{\beta N} \sum _{\omega _{m}{\bf k}} \left( \varepsilon_{\bf k} - \mu + \chi _{\bf k} \left( i\omega _{m}\right) \right) D _{\bf k}^{-1} \left( i\omega _{m}\right).
\end{equation}
\end{widetext}
The quantity $\varphi_{\bf k}\left(i\omega _{n}\right)$ is known as the order parameter function, and defined by the ratio:
$\Delta_{\bf k}\left(i\omega_{n}\right)=\varphi_{\bf k}\left(i\omega_{n}\right)/Z_{\bf k}\left(i\omega_{n}\right)$. In equations \eq{r1}-\eq{r3} the $K_{\bf kq}\left(\omega_{n}-\omega_{m}\right)$ function denotes the pairing kernel of the electron-phonon interaction: 
\begin{equation}
\label{r5}
K _{\bf kq} \left( \omega _{n}-\omega _{m}\right)=2g_{{\bf k}-{\bf q},{\bf k}}g_{{\bf k},{\bf k}+{\bf q}}\frac{\omega_{\qvec}}{\left( \omega _{n}-\omega _{m}\right)^{2}+\omega^{2}_{\qvec}},
\end{equation} 
whereas the phonon dispersion relation for the square lattice has the form: 
$\omega_{\qvec}=\omega_{0}\sqrt{2-\cos\left(q_{x}\right)-\cos\left(q_{y}\right)}$. 
In the present paper, we adopt $g_{{\bf k},{\bf k'}}\simeq g$. In the case of the square lattice, our assumption is justified by the results presented in \cite{Yonemitsu1989A}, where it is shown that the electron-phonon coupling function is characterized by the small anisotropy. Note that the relevance of the $g_{\kvec,\kvec'}$ anisotropy should be resolved individually for each analyzed superconducting material. For example, anisotropy of the electron-phonon interaction for the superconducting phase of niobium is negligible \cite{Butler1976A, Peter1977A}, whereas it has fundamental importance for the proper description of the superconducting state in magnesium diboride (${\rm MgB_{2}}$) \cite{Nagamatsu2001A, Golubov2002A, Choi2002A, Choi2003A}.
To this end, in the Eliashberg equations \eq{r1}-\eq{r3} the following substitution is introduced for the clarity of notation: 
$D_{\bf k}\left(i\omega_{n}\right)=\left(\omega_{n}Z_{\bf k}\left(i\omega _{n}\right)\right)^{2}+
\left(\varepsilon_{{\bf k}}-\mu+\chi_{{\bf k}}\left(i\omega_{n}\right)\right)^{2}+\left(\varphi_{\bf k}\left(i\omega_{n}\right)\right)^{2}$. 

The Eliashberg equations in the general form are too complex to be numerically solved at the sufficiently dense grid of the wave vector {\bf k} or {\bf q} (in our case $M\times M$, where $M=500$). The problem is that the number of the Eliashberg equations is equal to the number of the wave vector grid points multiplied by the Matsubara frequency number (typically several hundred frequencies \cite{Szczesniak2006A, Wiendloha2016A}). For this reason, we took into account Eliashberg equations independent of the Matsubara frequency, but we do not impose the specific dependence of the solutions on the wave vector. In particular, we use the following approximations: 
(i) $\varphi_{\kvec}\left(i\omega_{n}\right)\simeq \varphi_{\kvec}\left(i\omega_{n=1}\right)=\varphi_{\kvec}$, 
    $Z_{\kvec}\left(i\omega_{n}\right)\simeq Z_{\kvec}\left(i\omega_{n=1}\right)=Z_{\kvec}$, and 
    $\chi_{\kvec}\left(i\omega_{n}\right)\simeq \chi_{\kvec}\left(i\omega_{n=1}\right)=\chi_{\kvec}$.
(ii) We bring up the general Eliashberg equations to the form that the summation can be done:
\begin{eqnarray}
\label{r6}
\frac{1}{\beta}\sum_{\omega_{m}}\left[\frac{1}{E_{\kvec}-i\omega_{m}}
                         +\frac{1}{E_{\kvec}+i\omega_{m}}\right]
                         =\tanh\left(\frac{\beta E_{\kvec}}{2}\right).
\end{eqnarray}
If a given equation contains $\omega_{m}$ frequency term, it impossible to calculate the sum \eq{r6}, therefore we approximate it in the standard way: 
$\omega_{m}\simeq \omega_{m=1}=\pi/\beta$. 
(iii) In the same manner, we simplify the pairing kernel of the electron-phonon interaction: 
$K_{\bf kq}\left(\omega_{n}-\omega_{m}\right)\simeq 2g^{2}/\omega_{\qvec}$.  Note that the above approximations implies that the calculated critical temperature is higher than if we were determining $T_{C}$ with the help of the general Eliashberg equations \cite{Carbotte1990A}. Nonetheless, this fact does not affect the basic results presented in this paper. Finally, the Eliashberg equations take the form:
\begin{eqnarray}
\label{r7}
\varphi_{\kvec}=\frac{2g^{2}}{M}\sum_{\qvec}\frac{1}{\omega_{\qvec}}
\frac{\varphi_{\kvec-\qvec}}{Z^{2}_{\kvec-\qvec}}S\left(E_{\kvec-\qvec}\right),
\end{eqnarray}
\begin{eqnarray}
\label{r8}
Z_{\kvec}=
1+\gamma\frac{2g^{2}}{M}\sum_{\qvec}\frac{1}{\omega_{\qvec}Z_{\kvec-\qvec}}S\left(E_{\kvec-\qvec}\right),
\end{eqnarray}
\begin{eqnarray}
\label{r9}
\chi_{\kvec}=
-\gamma\frac{2g^{2}}{M}\sum_{\qvec}\frac{1}{\omega_{\qvec}}
\frac{\chi_{\kvec-\qvec}+\varepsilon_{\kvec-\qvec}-\mu}{Z^{2}_{\kvec-\qvec}}S\left(E_{\kvec-\qvec}\right),
\end{eqnarray}
\begin{eqnarray}
\label{r10}
\left<n\right>=1-\frac{2}{M}\sum_{\kvec}
\frac{\chi_{\kvec}+\varepsilon_{\kvec}-\mu}{Z^{2}_{\kvec}}S\left(E_{\kvec}\right),
\end{eqnarray}
where:
\begin{equation}
\label{r11}
E_{\kvec}=\sqrt{\left(\frac{\chi_{\kvec}+\varepsilon_{\kvec}-\mu}{Z_{\kvec}}\right)^{2}+\left(\frac{\varphi_{\kvec}}{Z_{\kvec}}\right)^{2}},
\end{equation}
and 
\begin{equation}
\label{r12}
S\left(E_{\kvec}\right)=\frac{\tanh\left(\frac{\beta E_{\kvec}}{2}\right)}{2E_{\kvec}}.
\end{equation}

In the present paper, we also consider the average values of the Eliashberg solutions: 
$\left<\Delta\right>=\frac{1}{M}\sum_{\kvec}(\varphi_{\kvec}\slash Z_{\kvec})$, 
$\left<Z\right>=\frac{1}{M}\sum_{\kvec}Z_{\kvec}$, and $\left<\chi\right>=\frac{1}{M}\sum_{\kvec}\chi_{\kvec}$.


\vspace*{0.25cm}
\begin{figure}
\includegraphics[width=\columnwidth]{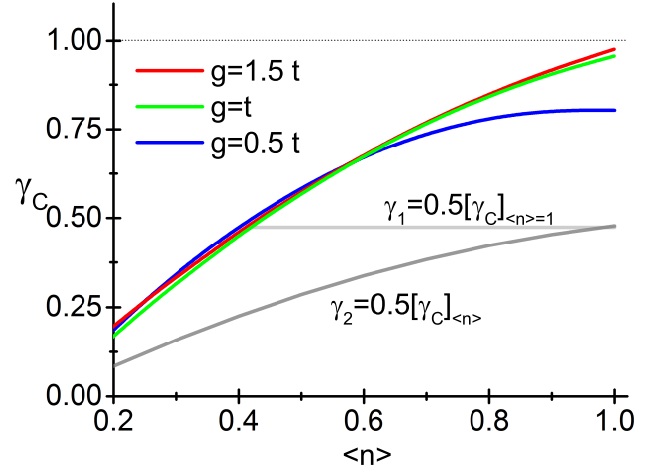}
\caption{The $\gamma_{C}$ dependence on the average number of electrons per site. 
         The selected values of $g$ and $T=T_{0}=8.62 \cdot 10^{-8}$~$t$ are adopted. 
         Two gray lines represent sets of $\gamma_{1}$ and $\gamma_{2}$ values, for which the thermodynamic properties 
         of the superconducting state are studied.}
\label{f01}
\end{figure} 
\begin{figure*}
\includegraphics[width=2.0\columnwidth]{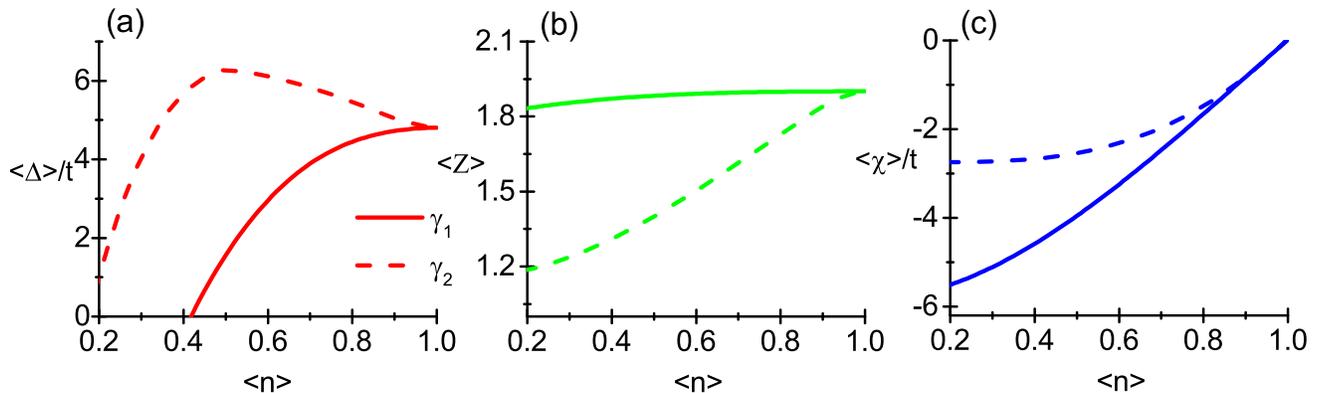}
\caption{(a) Average values of the order parameter, (b) the wave function renormalization factor, and (c) the band energy shift function as a function of the electron number density. The results are obtained for $\gamma=\gamma_{1}$ and $\gamma=\gamma_{2}$, when adopting that $g=t$ and $T=T_{0}$.}
\label{f02}
\end{figure*} 

\vspace*{0.25cm}

Numerical calculations are carried out assuming the following values of the input parameters: $\omega_{0}=0.15t$ and $g\in\{0.5t,t,1.5t\}$. \fig{f01} presents the most important result of the paper. Specifically, in the balanced case ($\gamma=1$), the linear electron-phonon interaction is not able to induce the superconducting state on the square lattice for any value of the electron number density. In particular, \fig{f01} shows the plot of the critical value of the unbalance parameter on $\left<n\right>$ for the minimum temperature $T_{0}$ equal to $8.62 \cdot 10^{-8}$~$t$. By the critical value of the unbalance parameter, we understand such value of $\gamma$ for which the superconducting state vanishes in the system (for $\gamma\geq\gamma_{C}$). On the basis of the presented data, it is true that this value increases with the electron number density, and reaches the maximum for the half-filled electron band. Nevertheless $\gamma_{C}$ is always less than one, and that is regardless of adopted $g$. Note that the disappearance of the superconducting state for $\gamma\geq \gamma_{C}$ is caused primarily by the too large electron effective mass 
($m^{\star}_{e}$) directly related to the value of the wave function renormalization factor: $m^{\star}_{e}=\left<Z\right>m_{e}$. It should be emphasized that the presented results can only be obtained when the Eliashberg equations are solved in the self-consistent manner in relation to ${\bf k}$. Note that when the self-consistent method of analysis is omitted it leads to the significant overestimation of $T_{C}$, which implies the induction of the superconducting state also in the balanced case (classic van Hove scenario) \cite{Goicochea1994A, Sarkar1995A, Markiewicz1997A, Mamedov1999A, Szczesniak2006B}.
\begin{figure*}
\includegraphics[width=2.0\columnwidth]{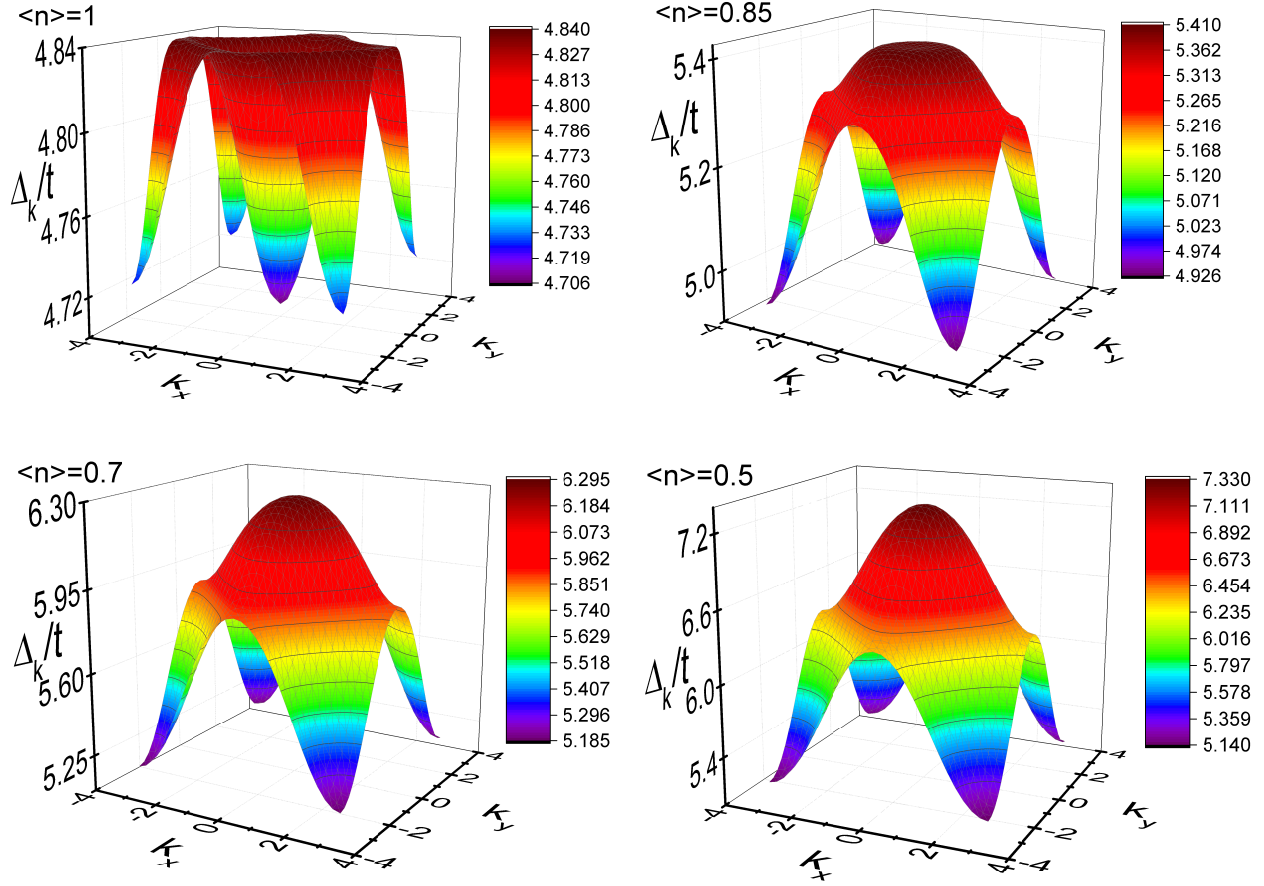}
\caption{The order parameter for the four selected values of $\left<n\right>$. It was adopted that $g=t$, $T=T_{0}$, and $\gamma=\gamma_{2}$. The equations were solved for the $M\times M$ lattice, where $M=500$. Note that the higher values of $M$ do not cause the change in the value of $\left<\Delta\right>$.}
\label{f03}
\end{figure*} 
\begin{figure*}
\includegraphics[width=\columnwidth]{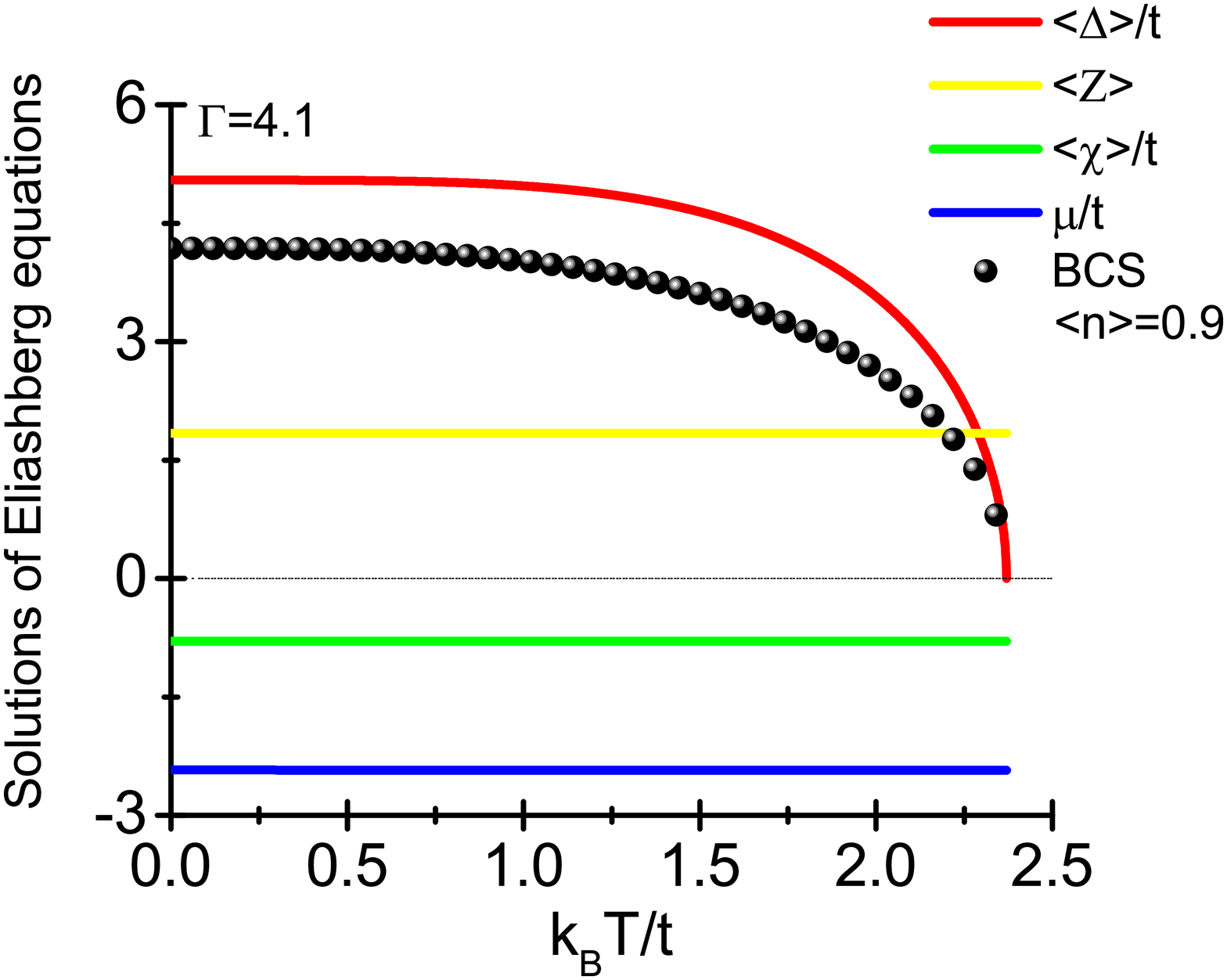}
\includegraphics[width=\columnwidth]{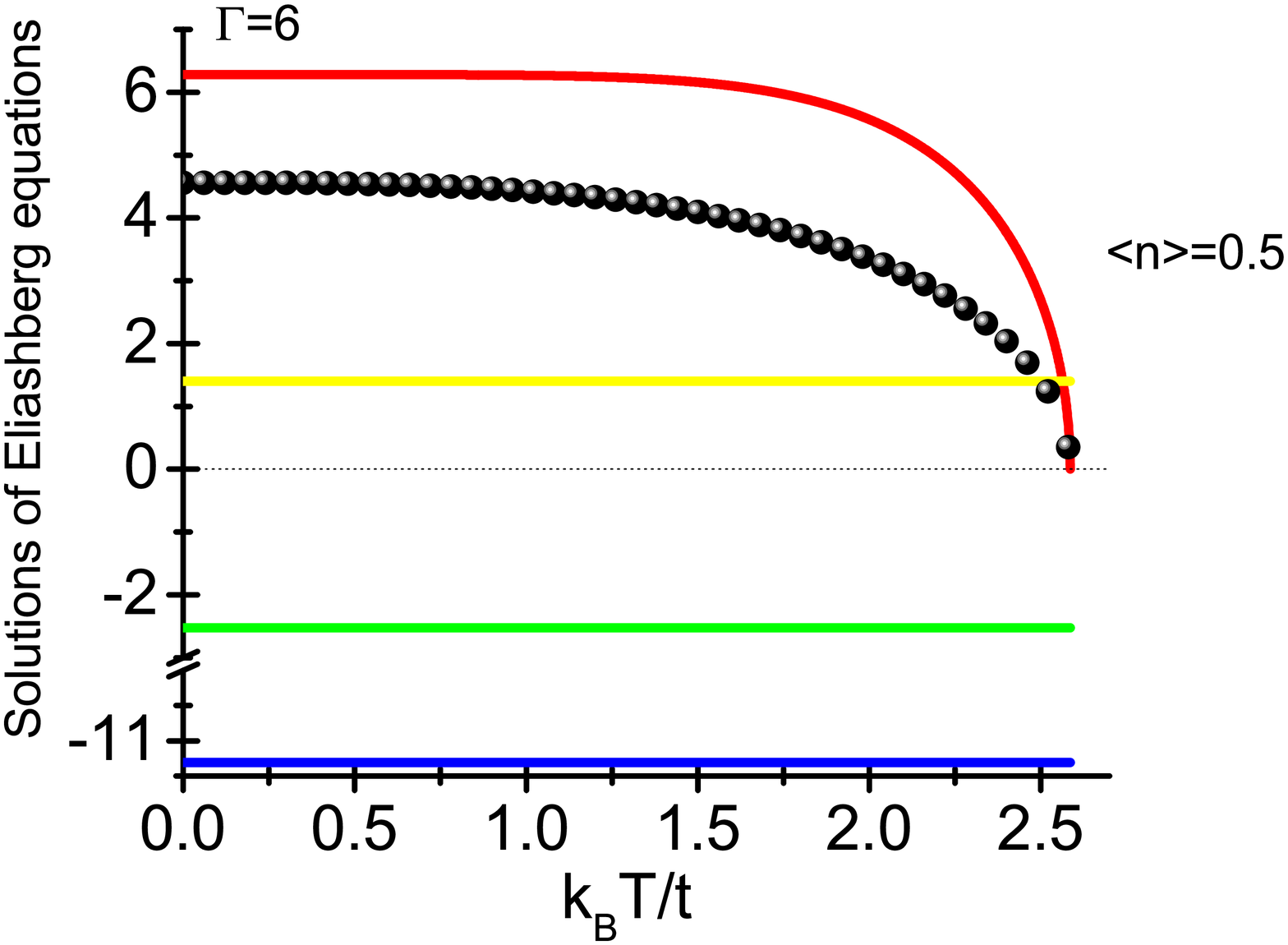}
\caption{The dependence of the averaged solutions of the Eliashberg equations on the temperature ($g=t$). The results were obtained within assumption: $\gamma=\gamma_{2}$, for $\left<n\right>=0.9$ and $\left<n\right>=0.5$. The black spheres correspond to the BCS theory \cite{Eschrig2001A}.}
\label{f04}
\end{figure*} 

Since the thermodynamic properties of the unbalanced superconducting state strongly depend on the adopted relation between the unbalance parameter and the electron number density, in the present paper, we consider the two most natural cases of such relation. At first, we assume that the unbalance parameter is equal to the half of $\gamma_{C}$ value for the half-filled electron band ($\gamma_{1}=0.5\left[\gamma_{C}\right]_{\left<n\right>=1}$). In the second case, we consider that the unbalance parameter depends on $\left<n\right>$ and is equal to the half of the value of $\gamma_{C}$, which corresponds to the given bandwidth ($\gamma_{2}=0.5\left[\gamma_{C}\right]_{\left<n\right>}$) - see also grey lines in \fig{f01}. The first case resembles the situation in which the unbalance is relatively strong, and is more destructive to the superconducting phase when the filling of electron band becomes smaller. It is worth noting that in the considered situation the superconducting state will disappear for $\left<n\right>=0.418$. In the second case, the unbalance parameter $\gamma_{2}$ is always below $\gamma_{C}$ and basically reproduces its dependence on $\left<n\right>$. 

\fig{f02} presents the influence of the electron number density on the solutions of the Eliashberg equations averaged over the wave vector. It can be observed that in the first case ($\gamma=\gamma_{1}$) the behavior of the thermodynamic functions is typical. In particular, the order parameter has the maximum for the half-filled electron band. On the other hand, the case $\gamma=\gamma_{2}$ gives completely non-standard dependence of the order parameter on $\left<n\right>$\textcolor{red}{,} with the maximum at $\left<n\right>\sim 0.5$. Physically this means that the explicit dependence of the unbalance parameter on the electron number density can lead to the non-classical phase diagram for the superconducting state.

The linear electron-phonon interaction cannot lead to the induction of the superconducting state with the strong anisotropy, {\it e.g.} $d$-wave \cite{Dagotto1994A}. Nevertheless, the dependence of the order parameter on the wave vector can be significant, causing the $\Delta_{\kvec}$ function to be inhomogeneous in the momentum space. This interesting fact was illustrated in \fig{f03} by plotting the exemplary values of the order parameter as a function of ${\bf k}$. It can be seen that the order parameter has particularly high heterogeneity for $\left<n\right>=0.5$. It should be also emphasized that functions $Z_{\kvec}$ and $\chi_{\kvec}$ are also heterogeneous. 

In the last step, we examined the dependence of the averaged solutions of the Eliashberg equations on the temperature. As expected, only the order parameter shows strong temperature dependence, what characterizes the phase transition of the second order. However, this dependence may differ from the predictions of the classical BCS theory (see \fig{f04}). Let us remind that in the mean-field BCS model, 
it was obtained that \cite{Eschrig2001A}:
\begin{equation}
\label{r13}
\Delta\left(T\right)=\Delta\left(0\right)\sqrt{1-\left(\frac{T}{T_{C}}\right)^{\Gamma}},
\end{equation}
while $R_{\Delta}=2\Delta\left(0\right)/k_{B}T_{C}=3.53$ and $\Gamma=3$. It is noteworthy that for the functional behavior presented in \fig{f04}~(a) the ratio $R_{\Delta}$ equals to $4.23$ and $\Gamma=4.1$, while the results from \fig{f04}~(b) give $R_{\Delta}=4.86$ and $\Gamma=6$. Increase in the value of the $R_{\Delta}$ parameter in relation to the canonical BCS value, associated with the reduced dimensionality of the tested system, equals $\sim 0.5$ at its maximum \cite{Szczesniak2002A}. Other contributions to $R_{\Delta}$ result from the strong-coupling effects included in the Eliashberg formalism \cite{Mitrovic1984A, Carbotte1990A, Carbotte2003A}.     


\vspace*{0.25cm}

In summary, we have shown that the linear electron-phonon interaction with the constant value of the coupling function is not able to induce the superconducting state on the square lattice. From the physical point of view, this fact occurs due to the excessive value of the electron effective mass. The obtained result does not mean that,during the analysis of the superconducting state in two-dimensional systems, the interaction of electrons and phonons can be omitted. On the contrary, it should be analyzed together with the other types of interactions ({\it e.g.} with the strong electron correlations \cite{Hubbard1963A}), which can lead to various coupling constants in the diagonal and non-diagonal channel of the self-energy. As a consequence, the unbalanced superconducting state may be induced, the properties of which will also depend on the electron-phonon coupling. However, in the case of the square lattice the thermodynamic parameters of the superconducting state are predominantly determined by the averaged electron-phonon coupling over the Fermi surface ($g_{\kvec,\kvec'}\simeq q$) \cite{Yonemitsu1989A}. We think that the present result open an interesting routes towards novel research on the high-$T_{C}$ superconducting materials like cuprates, where the total pairing, resulting from the two competing interactions (pure electron correlations and the electron-phonon coupling), is required to induce the unbalanced superconducting phase. 

Moreover, during presented analysis, we have shown that the explicit dependence of the unbalanced parameter on the electron number density may lead to the non-classical phase diagram $\Delta$-$\left<n\right>$. Next, based on the self-consistent analysis of the Eliashberg equations with respect to {\bf k}, we have shown that the functions $\Delta_{\kvec}$, $Z_{\kvec}$, and $\chi_{\kvec}$ can be characterized by the significant heterogeneity in the momentum space. This fact is often overlooked in the considerations related to the superconducting state induced by the linear electron-phonon interaction. In our opinion, it may be particularly important in the situations where the vertex corrections to the electron-phonon interaction are significant \cite{Pietronero1992A, Pietronero1995A, Grimaldi1995B, Grimaldi1995A}, because their importance is determined by the momentum dependence of $\Delta_{\kvec}$, $Z_{\kvec}$, and $\chi_{\kvec}$.

In the last part of the paper, we drew the reader's attention to the fact that the dependence of the order parameter on temperature for the unbalanced superconducting state may drastically differ from the predictions of classic BCS theory. This fact results from the reduced dimensionality of the considered system and the strong-coupling effects.  

\bibliography{bibliography}
\end{document}